\documentclass[conference]{IEEEtran}
\IEEEoverridecommandlockouts
\usepackage[mathscr]{eucal}
\usepackage[cmex10]{amsmath}
\usepackage{epsfig,epsf,psfrag}
\usepackage{amssymb,amsmath,amsthm,amsfonts,latexsym}
\usepackage{amsmath,graphicx,bm,xcolor,url,overpic}
\usepackage{array}
\usepackage{verbatim}
\usepackage{bm}
\usepackage{algorithm}
\usepackage{verbatim}
\usepackage{textcomp}
\usepackage{mathrsfs}
\usepackage{epstopdf}

\newcommand{\openone}{\leavevmode\hbox{\small1\normalsize\kern-.33em1}}

\catcode`~=11 \def\UrlSpecials{\do\~{\kern -.15em\lower .7ex\hbox{~}\kern .04em}} \catcode`~=13

\allowdisplaybreaks[4]

\newcommand{\nn}{\nonumber}

\newcommand{\calB}{\mathcal{B}}
\newcommand{\calC}{\mathcal{C}}
\newcommand{\calD}{\mathcal{D}}
\newcommand{\calE}{\mathcal{E}}
\newcommand{\calF}{\mathcal{F}}
\newcommand{\calG}{\mathcal{G}}
\newcommand{\calH}{\mathcal{H}}

\newcommand{\calM}{\mathcal{M}}

\newcommand{\calP}{\mathcal{P}}

\newcommand{\calR}{\mathcal{R}}
\newcommand{\calS}{\mathcal{S}}

\newcommand{\calX}{\mathcal{X}}

\newcommand{\by}{\mathbf{y}}
\newcommand{\bY}{\mathbf{Y}}

\newcommand{\rmA}{\mathrm{A}}

\newcommand{\rmE}{\mathrm{E}}

\newcommand{\rmH}{\mathrm{H}}

\newcommand{\rmN}{\mathrm{N}}

\newcommand{\rmP}{\mathrm{P}}

\newcommand{\rmr}{\mathrm{r}}

\newcommand{\bbE}{\mathsf{E}}

\newcommand{\bbN}{\mathbb{N}}

\newcommand{\bbP}{\mathbb{P}}

\newcommand{\bbR}{\mathbb{R}}

\DeclareMathAlphabet{\mathbsf}{OT1}{cmss}{bx}{n}
\DeclareMathAlphabet{\mathssf}{OT1}{cmss}{m}{sl}

\DeclareSymbolFont{bsfletters}{OT1}{cmss}{bx}{n}
\DeclareSymbolFont{ssfletters}{OT1}{cmss}{m}{n}
\DeclareMathSymbol{\bsfGamma}{0}{bsfletters}{'000}
\DeclareMathSymbol{\ssfGamma}{0}{ssfletters}{'000}
\DeclareMathSymbol{\bsfDelta}{0}{bsfletters}{'001}
\DeclareMathSymbol{\ssfDelta}{0}{ssfletters}{'001}
\DeclareMathSymbol{\bsfTheta}{0}{bsfletters}{'002}
\DeclareMathSymbol{\ssfTheta}{0}{ssfletters}{'002}
\DeclareMathSymbol{\bsfLambda}{0}{bsfletters}{'003}
\DeclareMathSymbol{\ssfLambda}{0}{ssfletters}{'003}
\DeclareMathSymbol{\bsfXi}{0}{bsfletters}{'004}
\DeclareMathSymbol{\ssfXi}{0}{ssfletters}{'004}
\DeclareMathSymbol{\bsfPi}{0}{bsfletters}{'005}
\DeclareMathSymbol{\ssfPi}{0}{ssfletters}{'005}
\DeclareMathSymbol{\bsfSigma}{0}{bsfletters}{'006}
\DeclareMathSymbol{\ssfSigma}{0}{ssfletters}{'006}
\DeclareMathSymbol{\bsfUpsilon}{0}{bsfletters}{'007}
\DeclareMathSymbol{\ssfUpsilon}{0}{ssfletters}{'007}
\DeclareMathSymbol{\bsfPhi}{0}{bsfletters}{'010}
\DeclareMathSymbol{\ssfPhi}{0}{ssfletters}{'010}
\DeclareMathSymbol{\bsfPsi}{0}{bsfletters}{'011}
\DeclareMathSymbol{\ssfPsi}{0}{ssfletters}{'011}
\DeclareMathSymbol{\bsfOmega}{0}{bsfletters}{'012}
\DeclareMathSymbol{\ssfOmega}{0}{ssfletters}{'012}

\newcommand{\tilx}{\tilde{x}}

\newcommand{\haty}{\hat{y}}

\newcommand{\tily}{\tilde{y}}

\newtheorem{theorem}{Theorem}
\newtheorem{lemma}{Lemma}

\usepackage{geometry}
\usepackage{graphicx,amssymb,amstext,amsmath}
\usepackage{makecell}
\usepackage{amsthm}
\usepackage{color}
\usepackage{subfigure}
\usepackage{algpseudocode}
\usepackage{multirow}
\usepackage[switch,pagewise]{lineno}
\usepackage{enumerate}
\usepackage{color}

\definecolor{Dyellow}{RGB}{254,152,0}
\definecolor{Dgreen}{RGB}{0,176,80}

\begin{document}
\title{Exponentially Consistent Low-Complexity Outlier Hypothesis Testing for Continuous Sequences\\
}
\author{  
\IEEEauthorblockN{Lina Zhu}
\IEEEauthorblockA{Space Information Research Institute\\
Hangzhou Dianzi University\\
Emails: zhulina@hdu.edu.cn}
\and
\IEEEauthorblockN{Lin Zhou}
\IEEEauthorblockA{School of Automation and Intelligent Manufacturing\\
Southern University of Science and Technology\\
Email: zhoul9@sustech.edu.cn}
}
\maketitle


\maketitle
\flushbottom

\begin{abstract}
In this work, we revisit outlier hypothesis testing and propose exponentially consistent, low-complexity fixed-length tests that achieve a better tradeoff between detection performance and computational complexity than existing exhaustive-search methods. In this setting, the goal is to identify outlying sequences from a set of observed sequences, where most sequences are i.i.d. from a nominal distribution and outliers are i.i.d. from a different anomalous distribution. While prior work has primarily focused on discrete-valued sequences, we extend the results of Bu et al. (TSP 2019) to continuous-valued sequences and develop a distribution-free test based on the MMD metric. Our framework handles both known and unknown numbers of outliers. In the unknown-count case, we bound the detection performance and characterize the tradeoff among the exponential decay rates of three types of error probabilities. Finally, we quantify the performance penalty incurred when the number of outliers is unknown.
\end{abstract}


\section{Introduction}
\label{sec:intro}



Outlier hypothesis testing (OHT) seeks to identify sequences that differ from the majority among a collection of $M$ sequences: most are i.i.d. from a nominal pdf $f_\rmN$ and the outliers are i.i.d. from an anomalous pdf $f_\rmA$. OHT has found extensive applications in a wide range of fields. Representative examples include detecting abnormal network traffic patterns to uncover sensitive information leakage from compromised computers \cite{kumar2005parallel}, and identifying anomalous magnetic resonance imaging (MRI) scans for early warning of malignant tumors \cite{spence2001detection}.

For discrete-valued observations, the performance of fixed-length OHT tests has been studied extensively. Li and Veeravalli~\cite{Universal_Outlier_Hypothesis}, and  Zhou, Wei, and Hero~\cite{Joint}, characterized the asymptotic performance limits under various settings. In the case of continuous-valued sequences, Zou \emph{et al.}~\cite{MMD} proposed an OHT framework based on the maximum mean discrepancy (MMD) metric defined in a reproducing kernel Hilbert space (RKHS)~\cite{Hilbert}. Building upon their work, we previously introduced a joint MMD-based detection scheme \cite{We}, which further improves the error performance by jointly exploiting pairwise discrepancies among sequences. For these methods, we analyzed the achievable exponential decay rates of the probabilities of misclassification, false rejection, and false alarm.

However, the tests proposed in \cite{Joint,We} rely on exhaustive search over all candidate outlier subsets, which leads to prohibitively high computational complexity. This severely limits their applicability in practical scenarios involving large numbers of sequences. To overcome this limitation, Bu, Zou, and Veeravalli~\cite{bu2019linear} proposed low-complexity testing procedures for discrete-valued sequences. They showed that their tests achieve exponential consistency when the number of outliers is known, and also proved exponential consistency when the number of outliers is unknown but strictly positive. However, analogous low-complexity and theoretically grounded solutions for continuous-valued sequences remain largely unexplored.

Motivated by this gap, we develop low-complexity, exponentially consistent tests for outlier hypothesis testing with continuous-valued sequences and analyze their large-deviations performance. When the number of outliers is known, we extend the approach of \cite{bu2019linear} to the continuous setting by proposing an MMD-based fixed-length test and deriving the exponential decay rate of the misclassification probability.

We further address the more general case of an unknown number of outliers, where misclassification, false rejection, and false alarm errors may occur. Removing the implicit assumption of a positive outlier count, we introduce an explicit detection phase and extend the framework of \cite{bu2019linear} to continuous observations. The resulting two-stage test first detects the presence of outliers via a thresholded maximum pairwise MMD statistic, and then identifies outliers using a simple thresholding rule that avoids exhaustive enumeration. This yields polynomial complexity in the number of sequences, independent of the true outlier count, and a favorable tradeoff between performance and complexity.

Finally, by comparing error exponents for known and unknown outlier counts, we quantify the performance loss due to uncertainty in the number of outliers.

\section{Basics}
\label{Problem_formulation}
\subsection*{Notation}
We use $\calR$, $\calR_+$, and $\bbN$ to denote the sets of real numbers, nonnegative real numbers, and natural numbers, respectively, and all logarithms are base $e$. Sets are denoted by calligraphic letters (e.g., $\calX$), random variables by uppercase letters (e.g., $X$), and their realizations by lowercase letters (e.g., $x$). For $n\in\bbN$, $Y^n=(Y_1,\ldots,Y_n)$ denotes a length-$n$ random vector. The set of all probability density functions on $\calR$ is denoted by $\calP(\calR)$. For $(a,b)\in\bbN^2$, $[a:b]$ denotes the set of integers from $a$ to $b$, and $[a]$ denotes $[1:a]$.

\subsection{Problem Formulation}

Fix $(n,M)\in\mathbb{N}^2$ and let ${f_\rmN,f_\rmA}\in\calP(\bbR)^2$ denote the nominal and anomalous distributions, respectively. In the framework of outlier hypothesis testing, one observes a collection of $M$ sequences $\bY^n={Y_1^n,\ldots,Y_M^n}$, each of length $n$. With the exception of a small, unknown subset, the sequences are composed of independent and identically distributed samples drawn from $f_\rmN$; the remaining sequences (if any) correspond to outliers and are generated i.i.d. according to $f_\rmA$. The objective is to devise a nonparametric testing rule that reliably identifies the outlier sequences, or correctly declares the absence of outliers.

\subsubsection{Case of Known Number of Outliers}

Fix $s\in\bbN$ such that $0<s\leq\lceil \frac{M}{2}-1\rceil$, and suppose that $s$ outliers exist among the $M$ observed sequences. Let 
$\calS_s=\{\calB\subseteq\calM:|\calB|=s\}$.
The goal is to construct a nonparametric decision rule
$\Phi_n:\bbR^{Mn}\to
\{\{\calH_\calB\}_{\calB \in\calS_s}\}$ 
that identifies which sequences are anomalous. Under hypothesis $\calH_\calB$, all sequences indexed by $j\in\calB$ are generated from the anomalous distribution and are therefore regarded as outliers, while the remaining sequences are nominal.
Fix any $\calB\in\calS_s$, and define
$ \calM_\calB:=[M]\setminus \calB = \{j\in [M] : j \notin \calB\}$.
To quantify the reliability of a test, we define the misclassification probability as:
\begin{align}
  \beta_\calB(\Phi_n|f_\rmN,f_\rmA)& := \rmP_\calB\{\Phi_n(\bY^n)\neq \rmH_\calB\}\label{beta_B_known}
\end{align}
where 
$\bbP_\calB(\cdot):=\rmP_\rmr\{\cdot|\rmH_\calB\}$  represents the probability measure induced by hypothesis $\rmH_\calB$.
Under $\rmH_\calB$, sequences indexed by $\calB$  are independently drawn from the anomalous distribution $f_\rmA$, while those indexed by 
$\calM_\calB$, are independently drawn from the nominal distribution $f_\rmN$.
Accordingly, 
$\beta_\calB(\cdot)$ captures the likelihood that the test fails to correctly recover the true set of outliers.

The following misclassification exponent describes the exponential decay rate of the misclassification probability.
\begin{align}
\rmE_\calB^\mathrm{Fix}(\Phi_n|f_\rmN,f_\rmA):=\liminf_{n\to\infty}
-\frac{1}{n}\log \beta_\calB(\Phi_n|f_\rmN,f_\rmA)  
\end{align}

\subsubsection{Case of Unknown Number of Outliers}

Let $T\in\bbN$ satisfy $0<T\leq\lceil \frac{M}{2}-1\rceil$, and suppose that the number of outliers is unknown but does not exceed $T$. Recall the collection of index sets $\calS_t$, and define $\calS:=\bigcup_{t\in[T]}\calS_t$. When the number of outliers is unknown, the objective is to construct a nonparametric decision rule that identifies the candidate outlier set while controlling false alarms. Equivalently, we seek a test  $\Phi_n:\bbR^{Mn}\to
\{\{\calH_\calB\}_{\calB \in\calS},\rmH_\rmr\}$ that discriminates among $|\calS|+1$ competing hypotheses.

\begin{itemize}
  \item $\rmH_\calB$,~$\calB\in \calS$: the sequences indexed by $\calB$ are outliers.
  \item $\rmH_\rmr$: no outlier is present.
\end{itemize}


For each $\calB\in\calS$, the performance of the proposed test is evaluated by the worst case misclassification and false reject
probabilities:
\begin{align}
\beta_\calB(\Phi_n|f_\rmN,f_\rmA)&:= \rmP_\rmr\{\Phi_n(\bY^n)\notin\{\rmH_\calB,\rmH_\rmr\}\},\label{Error_define2}\\  \zeta_\calB(\Phi_n|f_\rmN,f_\rmA)& := \rmP_\rmr\{\Phi_n(\bY^n)=\rmH_\rmr\},\label{Error_define3}
\end{align}
where $\bbP_\calB(\cdot)$ is defined similarly as in \eqref{beta_B_known}. The misclassification probability $\beta_\calB(\Phi_n \mid f_\rmN,f_\rmA)$ upper bounds the probability that the test $\Phi_n$ incorrectly identifies the set of outliers. The false reject probability $\zeta_\calB(\Phi_n \mid f_\rmN,f_\rmA)$ upper bounds the probability that the test $\Phi_n$ incorrectly declares the absence of outliers when outliers are in fact present.

Under the null hypothesis, the false alarm probability is defined as
\begin{align}
\rmP_{\rm{FA}}(\Phi_n \mid f_\rmN,f_\rmA)
&:= \bbP_\rmr\{\Phi_n(\bY^n)\neq \rmH_\rmr\},\label{FA}
\end{align}
where $\bbP_\rmr(\cdot) := \rmP_\rmr{\cdot \mid \rmH_\rmr}$  represents the probability measure induced by hypothesis $\rmH_\rmr$. 
Under the null hypothesis $\rmH_\rmr$, every sequence $Y_j^n$, $j\in[M]$, is independently drawn from the nominal distribution $f_\rmN$. Accordingly, the false alarm probability $\rmP_{\rm{FA}}(\cdot)$ measures the likelihood that the test $\Phi_n$ incorrectly reports the existence of outliers when all sequences are in fact nominal.

Similar to the case where the number of outliers is known, these exponents describe the  exponential decay behavior of the misclassification, false reject, and false alarm probabilities:
\begin{align}
\rmE_{\beta_\calB}(\Phi_n|f_\rmN,f_\rmA) &:=\liminf_{n\to\infty}-\log\frac{1}{n}\beta_\calB(\Phi_n|f_\rmN,f_\rmA)\\
\rmE_{\zeta_\calB}(\Phi_n|f_\rmN,f_\rmA) &:=\liminf_{n\to\infty}-\log\frac{1}{n}\zeta_\calB(\Phi_n|f_\rmN,f_\rmA)\\
\rmE_{\mathrm{FA}}(\Phi_n|f_\rmN,f_\rmA) &:=\liminf_{n\to\infty}-\log\frac{1}{n}\rmP_{\rm{FA}}(\Phi_n|f_\rmN,f_\rmA)
\end{align}

\subsection{MMD Metric}

Following \cite{MMD}, we employ the maximum mean discrepancy (MMD) metric \cite{gretton2012kernel} to construct our tests. For two distributions $f_1$ and $f_2$, the squared MMD is defined as
\begin{align}
\mathrm{MMD}^2(f_1,f_2)
:= \bbE_{f_1f_1}[k(X,X')] - &2\bbE_{f_1f_2}[k(X,Y)]\nn \\
&+ \bbE_{f_2f_2}[k(Y,Y')],
\end{align}
where $k(\cdot,\cdot)$ is a kernel associated with a reproducing kernel Hilbert space (RKHS), and $(X,X',Y,Y')\sim f_1f_1f_2f_2$. Clearly, $\mathrm{MMD}^2(f_1,f_2)=0$ if and only if $f_1=f_2$.

A commonly used choice is the Gaussian kernel
\begin{align}
k(x,y) := \exp\left(-\frac{(x-y)^2}{2\sigma_0^2}\right),\label{Gaussiankernel}
\end{align}
where $\sigma_0\in\bbR_+$ is a bandwidth parameter. We adopt this kernel in our numerical experiments.

Given samples $x^{n_1}=(x_1,\ldots,x_{n_1})$ and $y^{n_2}=(y_1,\ldots,y_{n_2})$ drawn i.i.d. from $f_1$ and $f_2$, respectively, the empirical squared MMD is
\begin{align}
\mathrm{MMD}^2(x^{n_1},y^{n_2})
\nn&:=\frac{1}{n_1(n_1-1)}\sum_{i,j\in[n_1],i\neq j}k(x_i,x_j)\\*
\nn&\qquad+\frac{1}{n_2(n_2-1)}\sum_{i,j\in[n_2],i\neq j}k(y_i,y_j)\\*
&\qquad-\frac{2}{n_1n_2}\sum_{i\in[n_1],j\in[n_2]}k(x_i,y_j)\label{MMDcompute}.
\end{align}
As shown in \cite[Lemma~6]{gretton2012kernel}, this statistic is an unbiased estimator of $\mathrm{MMD}^2(f_1,f_2)$ and converges almost surely to it as $n_1,n_2\to\infty$.

\section{Main Results}
\label{Main_unS}

\subsection{Main Results for the Known Number of Outliers}\label{S-FLMT}

\subsubsection{Test Design and Asymptotic Intuition}

We consider a fixed-length testing framework in which the observation length is a prescribed positive integer $n\in\bbN$. This section introduces a low-complexity testing procedure, detailed in Algorithm~\ref{LB_algorithm}. If the terminal \textbf{while} loop is omitted—so that the operations within the loop are performed only once—the resulting procedure is denoted by $\Phi_n(0)$ for a given sample size $n$.

\begin{algorithm}[ht!]
\caption{Low complexity algorithm with known number
$s$ of outliers}
\label{LB_algorithm}
\begin{algorithmic}[1]
\Require
$M$ observed sequences $\by^n=(\by_1^n,\ldots,\by_M^n)$ and the number $s$ of outliers;
\Ensure The set of outliers $\calB\in\calS_s$.
\State Select an index 
$j\in[M]$ uniformly at random and set the initial reference sequence as $y_0^n=y^n_j$.\label{FL_step0}
\State For each $i\in[M]$, compute $\mathrm{MMD}^2(y_0^n,y_i^n)$  
and arrange the resulting values in descending order to form the vector $\bf{v}_1$.
\State Let  $i^*$ denote the index corresponding to the $\lceil M/2\rceil$-th entry of $\mathbf{v}*1$, and set $\haty^n_\rmN\gets y^n_{i^*}$. \label{FL_step0-1}
\While{not converge}
   \State Recompute $\mathrm{MMD}^2(\haty_\rmN^n,y_i^n)$ for all $i\in[M]$ and sort the values in descending order to obtain $\bf{v}_2$.
   \State Define $\calB$ as the set of indices corresponding to the first $s$ entries of $\bf{v}_2$.\label{FL_step1}
   \State Update the estimate of the nominal sequence by
   \begin{align}\label{reEstimate}
  \haty^n_\rmN & \gets \arg\min_{y_i^n,i\in\calM_\calB}\mathrm{MMD}^2(y_i^n,\Bar{\bY}_{\calM_\calB,i}^n)
\end{align}
   where $\Bar{\bY}_{\calM_\calB,i}^n:=\{y_j^n|j\in \calM_\calB, j\neq i\}$ denotes the set of sequences in $\calM_\calB$ excluding $y_i^n$.\label{FL_step2}
\EndWhile
\State Return $\calB$.
\end{algorithmic}
\end{algorithm}

We next describe the key steps of our test. As shown in Algorithm~\ref{LB_algorithm}, from steps \ref{FL_step0}-\ref{FL_step0-1}, the test selects, with high probability, a sequence $\haty_\rmN^n$ whose generating distribution is close to the unknown nominal distribution $f_\rmN$. In the subsequent iterations, the test repeatedly identifies outliers via binary classification. In  step \ref{FL_step2}, the nominal reference sequence $\haty_\rmN^n$ is updated, and the procedure iterates until convergence, thereby mitigating errors caused by approximating the unknown generating distributions with empirical ones.

We now outline the asymptotic intuition that justifies the test’s performance. As the sample size $n$ grows, under hypothesis $\rmH_\calB$, the empirical distribution of each outlier sequence $y_i^n$ for $i\in\calB$ converges in probability to $f_\rmA$, while that of each nominal sequence $y_j^n$ for $j\in\calM_\calB$ converges to $f_\rmN$. Since the number of outliers satisfies $s< M/2$, nominal sequences form a strict majority. Consequently, with probability approaching one, the sequence $\haty_\rmN^n$ selected in the steps \ref{FL_step0}-\ref{FL_step0-1} is generated from $f_\rmN$.
Specifically, if the initial reference sequence $y_0^n$ is nominal, then for more than half of the sequences $y_j^n$, $j\in\calM_\calB$, the pairwise MMD satisfies $\mathrm{MMD}^2(y_j^n,y_0^n)\to 0$, implying $\mathrm{MMD}^2(\haty_\rmN^n,y_0^n)\to 0$. If $y_0^n$ is an outlier, then for a majority of nominal sequences the MMD converges to $\mathrm{MMD}(f_\rmN,f_\rmA)$, again leading to the correct selection of a nominal reference.
With $\haty_\rmN^n$ correctly identified, the assignment step succeeds: for nominal sequences $y_i^n$, $i\in\calM_\calB$, $\mathrm{MMD}^2(\haty_\rmN^n,y_i^n)\to 0$, while for outliers $y_j^n$, $j\in\calB$, $\mathrm{MMD}^2(\haty_\rmN^n,y_j^n)\to \mathrm{MMD}(f_\rmN,f_\rmA)$, enabling reliable separation of outliers from nominal samples.

%

\subsubsection{Theoretical Results and Discussions}

In the following discussions, let $\beta_\calB(\Phi_n(l)|f_\rmN,f_\rmA)$ be the misclassification probability after $l$ rounds of iteration. 

\begin{theorem}\label{FL}
For any choice of distributions $(f_\rmN,f_\rmA)\in\calP(\bbR)^2$ and for every candidate outliers set $\calB\in\calS_s$,  misclassification error exponent of the fixed-length testing procedure described in Algorithm~\ref{LB_algorithm} satisfies
 \begin{align}
  \rmE_\calB^\mathrm{Fix}(\Phi_n(0)|f_\rmN,f_\rmA)\geq \frac{\mathrm{MMD}^4(f_\rmN,f_\rmA)}{96K_0^2}.\label{Exp_known}
    \end{align}
    When $s<\frac{M}{3}$, it is satisfied that
    \begin{align}
        \rmE_\calB^\mathrm{Fix}(\Phi_n(l)|f_\rmN,f_\rmA)>  \rmE_\calB^\mathrm{Fix}(\Phi_n(0)|f_\rmN,f_\rmA).
    \end{align}
for $l\to\infty$.

\end{theorem}

The proof of Theorem \ref{FL} is provided in Appendix \ref{proof_of_FL}.
Theorem~\ref{FL} establishes that the misclassification exponent of our low-complexity test in Algorithm~\ref{LB_algorithm} is lower bounded by the divergence between the nominal and anomalous distributions, $f_\rmN$ and $f_\rmA$. Specifically, the dominant error events arise when either the initially selected center $\hat{y}_\rmN^n$ is an outlier sequence, or the assignment step incorrectly labels a nominal sequence as an outlier.
As discussed in Appendix \ref{proof_of_FL}, $\calE_{\mathcal{B},1}$ occurs when the inter-cluster MMD between two sequences drawn from different distribution clusters is smaller than the intra-cluster MMD computed within the normal distribution cluster.  $\calE_{\mathcal{B},2}$ occurs when the $\lceil \frac{M}{2}\rceil$-th element of $\bf{v}_1$ corresponds to the MMD metric between a nominal sequence and an outlier.  The error event during the step \ref{FL_step0}-\ref{FL_step0-1} includes the cases of both $\calE_{\mathcal{B},1}$ and $\calE_{\mathcal{B},2}$, while the error event during the assignment step only includes $\calE_{\mathcal{B},1}$. The proof of Theorem \ref{FL} shows that the exponential
decay rate for the probabilities of both $\calE_{\mathcal{B},1}$ and $\calE_{\mathcal{B},2}$ are lower bounded by $\frac{\mathrm{MMD}^4(f_\rmN,f_\rmA)}{96K_0^2}$.

For the case of $l>0$ iterations, we analyze the error probability in the $r$-th iteration for $r\in[l]$. Let $\hat{y}\rmN^n$ be the re-estimated nominal center at round $r-1$ according to \eqref{reEstimate}. If $\hat{y}\rmN^n$ at round $r-1$ is actually an outlier, then the probability of error at round $r$ approaches $1$. Consequently, the error event at iteration $r$ occurs under either of the following conditions:
i) $\hat{y}_\rmN^n$ at round $r-1$ is an outlier, or
ii) A nominal sequence is misclassified as an outlier during Step~\ref{FL_step1} at round $r$.
Let $\calE_\mathrm{error}^{(r)}$ denote the event of misclassification at round $r$, and let $\calE_{\mathcal{B},\mathrm{A}}^{(r)}$ denote the event that $\hat{y}\rmN^n$ is an outlier after $r$ rounds. Then, $\mathbb{P}_\calB\{\calE_{\mathcal{B},\mathrm{A}}^{(r)} \mid \calE_\mathrm{error}^{(r)}\}$ represents the probability that the selected nominal center is actually an outlier at round $r$, conditional on a misclassification occurring. We show that this probability decays exponentially when $s < M/3$, which establishes the main claim of Theorem~\ref{FL}.

Theorem~\ref{FL} indicates that when $s < M/3$, the test can leverage the performance gains provided by multiple iterations.
Note that in other cases, $ \rmE_\calB^\mathrm{Fix}(\Phi_n(l)|f_\rmN,f_\rmA)>  \rmE_\calB^\mathrm{Fix}(\Phi_n(0)|f_\rmN,f_\rmA)$ in Theorem \ref{FL} is not always satisfied for any distribution $f_\rmN,f_\rmA$. This is because we cannot guarantee $\bbP_\calB\{\calE_{\calB,\rmA}^{(r)}|\calE_\mathrm{error}^{(r)}\}$ is exponential consistent. Specifically, we have
\begin{align}
&\beta_\calB(\Phi_n(l)|f_\rmN,f_\rmA)\nn\\
&\leq (\bbP_\calB\{\calE_{\calB,1}\}+\bbP_\calB\{\calE_{\calB,2}\})\prod_{r\in[0:l-1]}\bbP_\calB
\{\calE_{\calB,\rmA}^{(r)}|\calE_\mathrm{error}^{(r)}\}\nn\\
&+\bbP_\calB(\calE_{\calB,3})\left(1+\sum_{i\in[0:l-1]}\prod_{r=l-1-i}^{l-1}\bbP_\calB\{\calE_{\calB,\rmA}^{(r}|\calE_\mathrm{error}^{(r)}\}\right)\\
&\leq \bbP_\calB\{\calE_{\calB,1}\}+\bbP_\calB\{\calE_{\calB,2}\}+(1+l)\bbP_\calB\{\calE_{\calB,3}\}\label{betaB2}.
\end{align}
The exponent of \eqref{betaB2} is lower bounded by \eqref{error_known}, which indicates that the misclassification probability is largely determined by errors occurring in the first iteration, and additional iterations offer negligible improvement in performance. 
This observation highlights that the benefit of iterative refinement is significant only when the number of outliers is sufficiently small relative to the total number of sequences.



\subsection{Main Results for the Unknown Number of Outliers}
\label{S-FL-un}

\subsubsection{Test Design and Asymptotic Intuition}

This subsection presents our low-complexity fixed-length test when
the number of outliers is unknown.

Fix $\lambda\in\bbR_+$. The proposed test, summarized in Algorithm~\ref{unknown_algorithm}, operates in two phases: outlier detection (Steps~\ref{UNFx_1}–\ref{UNFx_2}) and outlier identification. In the detection phase, the test computes all pairwise MMD statistics and declares the absence of outliers if the maximum MMD is below the threshold $\lambda$; otherwise, it proceeds to outlier identification.
In the identification phase, two cluster centers are selected: $c_1$ is chosen randomly, and $c_2$ is chosen as the sequence with the largest MMD relative to $c_1$. The test then performs binary clustering using the minimum-MMD decision rule to form two clusters, $\calC_1$ and $\calC_2$, and declares the smaller cluster as the set of outliers.

\begin{algorithm}[ht!]
\caption{Low complexity algorithm with unknown number
of outliers}
\label{unknown_algorithm}
\begin{algorithmic}[1]
\Require
$M$ observed sequences $\by^n$ and a positive threshold $\lambda\in\bbR_+$
\Ensure A hypothesis in the set $\{\{\rmH_\calB\}_{\calB\in\calS},\rmH_\rmr\}$
\State Compute $\mathrm{MMD}^2(y_i^n,y_j^n)$ for all $(i,j)\in[M]^2$\label{UNFx_1}
\If{$\max_{i,j\in[M]^2,i\neq j}\mathrm{MMD}^2(y_i^n,y_j^n)<\lambda$}
   \State Return Hypothesis $\rmH_\rmr$.\label{UNFx_2}
\Else
\State Choose a number $i \in[M]$ randomly
\State  Calculate $j^*:=\arg\max_{j\in[M]}\mathrm{MMD}^2(y_i^n,y_j^n)$
\State Set $c_1=y_i^n$ and $c_2=y_{j^*}^n$
\State Set $\calC_1\gets\emptyset$, $\calC_2\gets\emptyset$;
  \For{$i\in [M]$}
  \State Calculate $j^*=\arg\min_{j\in[2]}\mathrm{MMD}^2(c_j,y_i^n)$
  \State $\calC_{j^*}\gets \calC_{j^*}\cup \{y_i^n\}$;
  \EndFor
  \State Calculate $t^*= \arg \min_{t\in[2]} |\calC_t|$
  \State Return Hypothesis $\rmH_{\calC_{t^*}}$.
\EndIf
\end{algorithmic}
\end{algorithm}

We now provide asymptotic intuition for the correctness of the proposed test. As $n\to\infty$, the empirical statistic $\mathrm{MMD}^2(y_i^n,y_j^n)$ converges to its expectation, which equals zero when both sequences are generated from the same distribution (both nominal or both outliers), and converges to $\mathrm{MMD}^2(f_\rmN,f_\rmA)$ when one sequence is nominal and the other is an outlier.
In the outlier detection phase, if no outliers are present, all pairwise MMD values converge to zero, and the test correctly declares $\rmH_\rmr$ for any threshold $0<\lambda<\mathrm{MMD}^2(f_\rmN,f_\rmA)$. Conversely, if outliers exist, there are indices $i\in\calB$ and $j\in\calM_\calB$ such that $\mathrm{MMD}^2(y_i^n,y_j^n)$ exceeds $\lambda$, triggering the outlier identification phase.
In the identification phase, with probability approaching one, the two cluster centers $c_1$ and $c_2$ correspond to a nominal sequence and an outlier, although their order is arbitrary. Accordingly, the resulting clusters $\calC_1$ and $\calC_2$ asymptotically separate nominal sequences from outliers. Since the number of outliers is strictly smaller than the number of nominal sequences, the cluster with fewer elements is correctly identified as the outlier set.

\subsubsection{Theoretical Results and Discussions}


\begin{theorem}\label{FL_un}
Fix any $\lambda\in\bbR_+$, under any pair of distributions $(f_\rmN,f_\rmA)\in\calP(\bbR)^2$

\begin{itemize}
    \item for each $\calB\in\calS$,
    \begin{itemize}
      \item the misclassification exponent satisfies
 \begin{align}
     \rmE_{\beta_\calB}(\Phi_n|f_\rmN,f_\rmA) \geq \frac{\mathrm{MMD}^4(f_\rmN,f_\rmA)}{96K_0^2}\label{mis_error_exponent_unknown}
 \end{align}
        
        \item the false reject exponents satisfies
\begin{align}
   \rmE_{\zeta_\calB}(\Phi_n|f_\rmN,f_\rmA) \geq \frac{(\mathrm{MMD}^2(f_\rmN,f_\rmA)-\lambda)^2}{64K_0^2}
\end{align}   
    \end{itemize}

\item
the false alarm exponent satisfies
\begin{align}   \rmE_{\mathrm{FA}}(\Phi_n|f_\rmN) \geq \frac{\lambda^2}{64K_0^2}
\end{align}
\end{itemize}

\end{theorem}

The proof of Theorem \ref{FL_un} has been given in Appendix \ref{proof_of_FL_un}.
As established in Theorem~\ref{FL_un}, the misclassification corresponds to two dominant error events: i) both cluster centers $c_1$ and $c_2$ are either outliers or nominal samples, and ii) an outlier is misclassified as nominal or a nominal sample is misclassified as an outlier. These two events are shown to yield the same error exponent in the limit as $n\to\infty$.
In addition, the false reject and false alarm exponents depend critically on the threshold $\lambda$. Specifically, $\lambda$ governs a tradeoff between these two errors: the false reject exponent $\frac{(\mathrm{MMD}^2(f_\rmN,f_\rmA)-\lambda)^2}{64K_0^2}$ is non-increasing in $\lambda$, whereas the false alarm exponent $\frac{\lambda^2}{64K_0^2}$ is non-decreasing in $\lambda$. Moreover, the false alarm exponent is strictly positive for all $\lambda\in\bbR_+$, while the false reject exponent is strictly positive whenever $\lambda<\mathrm{MMD}^2(f_\rmN,f_\rmA)$.

By comparing Theorems~\ref{FL} and~\ref{FL_un}, we quantify the performance penalty incurred by not knowing the number of outliers in low-complexity fixed-length tests under non-null hypotheses. In Theorem~\ref{FL}, the number of outliers $s$ is assumed known, whereas in Theorem~\ref{FL_un} it is unknown but upper bounded by $T$. For a fair comparison, we evaluate the error probability under each non-null hypothesis. Accordingly, we compare the misclassification exponent in Theorem~\ref{FL}, $\rmE_{\beta_\calB}^\mathrm{Fix}(\Phi_n\mid f_\rmN,f_\rmA)$, with the minimum of the misclassification and false reject exponents in Theorem~\ref{FL_un}, namely $\min\{\rmE_{\beta_\calB}(\Phi_n\mid f_\rmN,f_\rmA),\rmE_{\zeta_\calB}(\Phi_n\mid f_\rmN,f_\rmA)\}$.
 It follows that 
\begin{align}
  &\frac{\mathrm{MMD}^4(f_\rmN,f_\rmA)}{96K_0^2}\nn\\ 
  &\geq \min \bigg\{\frac{\mathrm{MMD}^4(f_\rmN,f_\rmA)}{96K_0^2},\frac{(\mathrm{MMD}^2(f_\rmN,f_\rmA)-\lambda)^2}{64K_0^2}\bigg\},
\end{align}
where the equalization is satisfied if and only if $0<\lambda\leq (1-\sqrt{2/3})\mathrm{MMD}^4(f_\rmN,f_\rmA)$.
Thus, the fixed-length test that knows the number of outliers has better performance than the fixed-length test that
does not know the number of outliers if $\lambda> (1-\sqrt{2/3})\mathrm{MMD}^4(f_\rmN,f_\rmA)\sim 0.18\mathrm{MMD}^4(f_\rmN,f_\rmA)$.

\section{Numerical Results}\label{simulation}

In this section, we evaluate the performance of our proposed test and compare it with the tests in \cite[Eq.~(13)]{MMD} and \cite{We}. Without loss of generality, we assume the first $s$ sequences among $M$ observed sequences $\bY^n=(Y_1^n,\dots,Y_M^n)$ are outliers. We set $f_\rmN=\mathcal{N}(0,1)$ and $f_\rmA=\mathcal{N}(1.5,1)$, i.e., Gaussian distributions with equal variance and different means. For the MMD metric~\eqref{MMDcompute}, we use the Gaussian kernel~\eqref{Gaussiankernel} with $\sigma=1$, and set $M=10$ unless otherwise stated.

\begin{figure}
    \centering
    \includegraphics[width=0.8\linewidth]{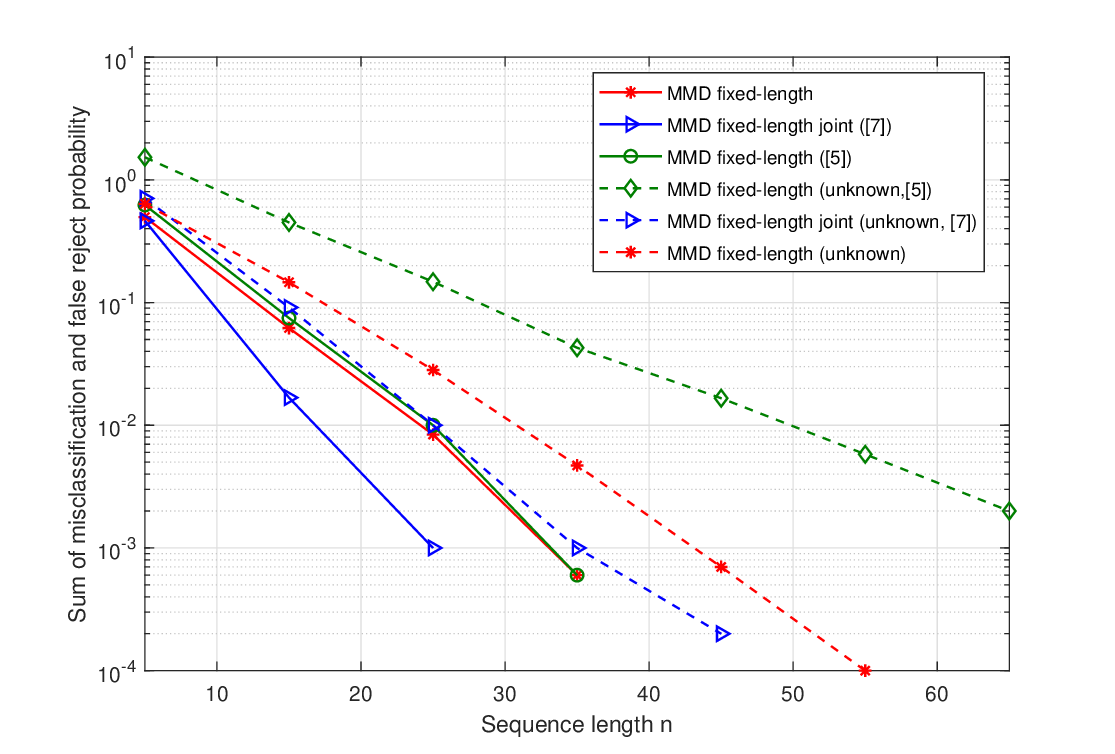}
    \caption{ Plot of the sum of simulated misclassification and the false reject probabilities as a function of sequence length for the test in Algorithm \ref{LB_algorithm} and \ref{unknown_algorithm}, the joint MMD tests proposed in \cite{We} and the tests in \cite{MMD} when $M = 10$, $s=2$, the threshold $\lambda$ in the unknown case is defined by $\lambda=0.3\mathrm{MMD}^2(f_\rmN,f_\rmA)$. }
    \label{fig_s2}
\end{figure}

Fig.~\ref{fig_s2} shows the sum of simulated error probabilities for the fixed-length tests in Algorithms~\ref{LB_algorithm} and~\ref{unknown_algorithm}, when $s=2$ outliers are present among $M=10$ sequences. The sample length is varied over $n\in[65]$. For the unknown case, the number of outliers is bounded by $T\le \lceil M/2\rceil-1 = 4$. Error probabilities are averaged over 10,000 independent runs.
As seen in Fig.~\ref{fig_s2}, for unknown outlier counts, our low-complexity test outperforms the test in \cite{MMD}, though it remains inferior to the test in \cite{We}, reflecting the tradeoff for reduced complexity. Comparing the performance of Algorithms~\ref{LB_algorithm} and~\ref{unknown_algorithm} also confirms the performance penalty incurred when the number of outliers is unknown, consistent with the discussion following Theorem~\ref{FL_un}.

\begin{figure}
    \centering
    \includegraphics[width=0.8\linewidth]{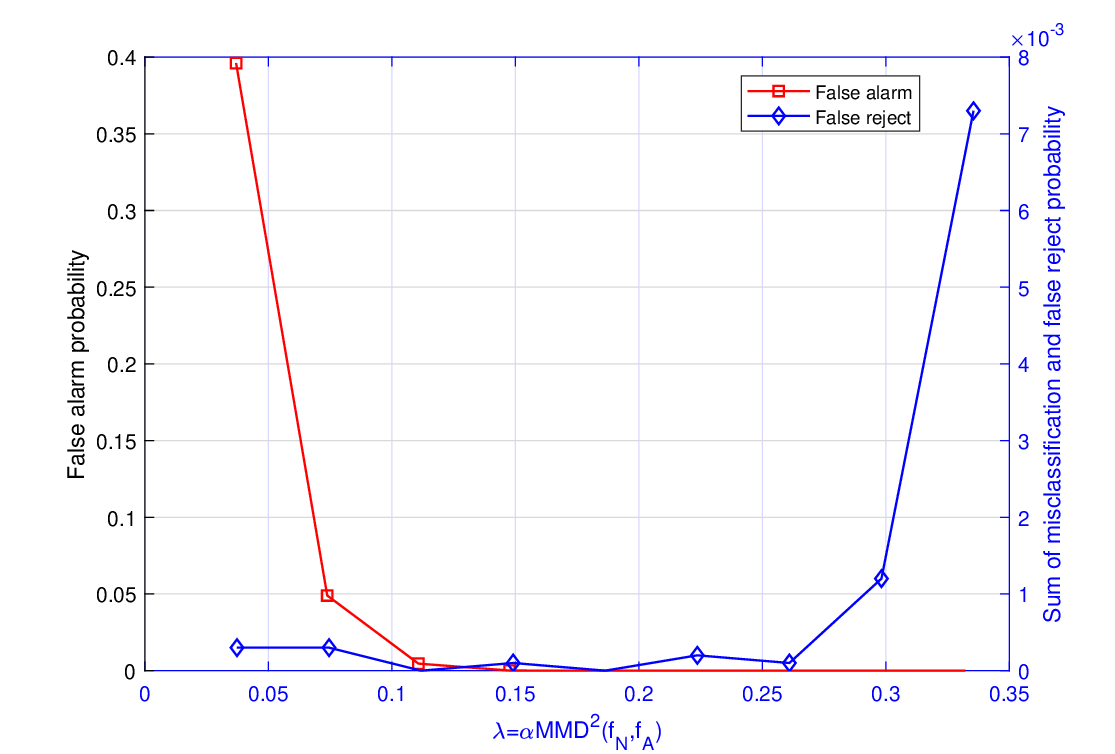}
    \caption{ Plot of the error probability under each non-null hypothesis and null hypothesis as a function of $\lambda$ for the test in Algorithm  \ref{unknown_algorithm} when $M = 10$, $n=60$,$s=2$. The threshold $\lambda$ in the unknown case is defined by $\lambda=\alpha\mathrm{MMD}^2(f_\rmN,f_\rmA)$, where $\alpha$ is selected from $0.1$ to $0.9$.  }
    \label{lambda}
\end{figure}

Fig.~\ref{lambda} shows the error probabilities of Algorithm~\ref{unknown_algorithm} under both the null and each non-null hypothesis as a function of $\lambda$, for $s=2$ outliers among $M=10$ sequences with sample length $n=60$. Probabilities are averaged over 10,000 independent runs.
As the figure illustrates, when the number of outliers is unknown, the threshold $\lambda$ governs the tradeoff between errors under non-null and null hypotheses. Larger $\lambda$ improves the error probability under non-null hypotheses while reducing the false alarm probability. Moreover, under a non-null hypothesis, the error probabilities of Algorithm~\ref{unknown_algorithm} can approach those of the known-outlier case when $0<\lambda \le (1-\sqrt{2/3})\mathrm{MMD}^4(f_\rmN,f_\rmA)$. However, choosing too small a $\lambda$ leads to excessively high false alarm probability.

\section{Conclusion}
\label{sec:conc}
This work revisits the problem of outlier hypothesis testing and develops a class of fixed-length tests that are both exponentially consistent and computationally efficient, without requiring prior knowledge of the nominal or anomalous distributions. The proposed framework accommodates scenarios in which the number of outliers is either known or unknown. We rigorously analyze the detection performance by deriving the achievable error exponents associated with multiple types of decision errors. A key advantage of the proposed methods is that their computational complexity scales only polynomially with the total number of observed sequences and remains independent of the number of outliers. In contrast to existing approaches such as \cite{We}, which rely on exhaustive search and thus suffer from prohibitive computational costs, our tests achieve a more favorable balance between statistical performance and algorithmic efficiency. Moreover, by contrasting the cases of known and unknown outlier counts, we explicitly quantify the performance degradation incurred when the number of outliers is not available a priori, for both fixed-length and sequential testing paradigms.

We next discuss several future research directions. First, it would be of interest to extend our fixed-length tests to the sequential testing framework \cite{We}. Second, our analysis assumes that all nominal sequences are generated from the same nominal distribution and all outliers from a single anomalous distribution. In practice, however, nominal samples may follow slightly different distributions, and the same holds for outliers. To better capture such practical scenarios, it is worthwhile to generalize our results to settings with distributional uncertainty, for example by using exponential family models \cite{pan2022asymptotics} or distributional uncertainty sets such as distribution balls \cite{binaryI-Hsiang,huber1965robust}. Finally, it would be of great interest to extend the low-complexity test construction techniques developed in this paper to other statistical inference problems, including distributed detection \cite{tenney1981detection,tsitsiklis1988decentralized}, quickest change-point detection \cite{poor2008quickest,tartakovsky2014sequential}, and clustering \cite{kaufman1990finding,park2009simple}.



\newpage
\bibliographystyle{IEEEtran}
\bibliography{BiB}
\newpage
\appendix

\subsection{Proof of Theorem \ref{FL}}\label{proof_of_FL}

Consistent with \cite{MMD}, we need the following McDiarmid's inequality \cite{mcdiarmid1989method} to bound error probabilities.
\begin{lemma}
\label{McDiarmid}
Let $g:\calX^n\rightarrow \mathbb{R}$ be a
function such that for each $k\in[n]$, there exists a constant $c_k<\infty$ such that
\begin{align}\label{McDiarmid1}
\nn\sup_{x^n\in\calX^n,\tilx\in\calX}&\big|g(x_1,\ldots,x_k,\ldots,x_n)\\
&-g(x_1,\ldots, x_{k-1},\tilx,x_{k+1},\ldots,x_n)\big|\leq c_k.
\end{align}
Let $X^n$ be generated i.i.d. from a pdf $f\in\calP(\bbR)$. For any $\varepsilon>0$, it follows that
\begin{align}
\label{McDiarmid2}
\mathrm{Pr}\big\{g(X^n)-\bbE_f[g(X^n)]>\varepsilon\big\}< \exp\left\{-\frac{2\varepsilon^2}{\sum_{i=1}^nc_i^2}\right\}.
\end{align}
\end{lemma}

We first consider the case of $l=0$.
As shown in Algorithm \eqref{LB_algorithm}, the misclassification error event may occur in the following two cases:
\begin{itemize}
\item $\haty_\rmN^n$ is actually generated from an outlying distribution. 
\item Given that $\haty_\rmN^n$ is actually
generated from nominal distribution, the generated
distribution of an outlying sequence is closer to anomalous distribution sets.
\end{itemize}

For the first case, if $y_0^n$ is generated
from the nominal distribution, an error occurs when $\haty_\rmN^n$ is actually the outlying sequence,  then there must exist $i\in\calB$, $j\in\calM_\calB$ that make $\mathrm{MMD}^2(y_i^n,y_0^n)<\mathrm{MMD}^2(y_j^n,y_0^n)$. In this case, the probability
of this error event can be upper-bounded by the probability of
the following event:
\begin{align}
\calE_{\calB,1}:~\exists~i\in \calB, &\exists~j_1,j_2\in\calM_\calB,\nn\\
&\mathrm{MMD}^2(y_i^n,y_{j_1}^n)<\mathrm{MMD}^2(y_{j_2}^n,y_{j_1}^n)\label{EB1}
\end{align}
On the other hand, if $y_0^n$ is actually an outlier, the error
probability can be upper bounded by the probability of the following
event:
\begin{align}
\calE_{\calB,2}:&~\exists~i_1,i_2\in \calB,\exists~j_1,j_2\in\calM_\calB,\nn\\
&\mathrm{MMD}^2(y_{j_1}^n,y_{i_1}^n)<\mathrm{MMD}^2(y_{j_2}^n,y_{j_1}^n)<\mathrm{MMD}^2(y_{j_2}^n,y_{i_1}^n)\label{EB2}
\end{align}

For the second case, when $\haty_\rmN^n$ is correctly selected, the error occurs when the following event happens:
\begin{align}
& \calE_{\calB,3}:~\exists~i\in \calB, \exists~j\in\calM_\calB,\mathrm{MMD}^2(y_i^n,\haty_\rmN^n)<\mathrm{MMD}^2(y_j^n,\haty_\rmN^n)\label{EB3}
\end{align}
Therefore, $\beta_\calB(\Phi_n(0)|f_\rmN,f_\rmA)$ can be upper bounded by
\begin{align}
&\beta_\calB(\Phi_n(0)|f_\rmN,f_\rmA) \\
&=\bbP_\calB\{\calE_{\calB,1}\cup \calE_{\calB,2}\}+\bbP_\calB\{(\calE_{\calB,1}\cup \calE_{\calB,2})^c\}\bbP\{\calE_{\calB,3}\}\\
&\leq \bbP_\calB\{\calE_{\calB,1}\}+\bbP\{\calE_{\calB,2}\}+\bbP_\calB\{\calE_{\calB,3}\}\\
&\leq 3\sum_{\substack{i\in\calB,\\j_1,j_2\in\calM_\calB,\\j_1\neq j_2}}\bbP_\calB\{\mathrm{MMD}^2(Y_i^n,Y_{j_1}^n)<\mathrm{MMD}^2(Y_{j_2}^n,Y_{j_1}^n)\}\label{error_known}
\end{align}

We apply McDiarmid's inequality to further upper bound \eqref{error_known}. To do so, with the result of \eqref{McDiarmid2},  we need  to calculate the expected value of $\mathrm{MMD}^2(Y_{j_2}^n,Y_{j_1}^n)-\mathrm{MMD}^2(Y_i^n,Y_{j_1}^n)$ and the parameters $c_k$ for each $k\in[3n]$, $i\in\calB$, and $j_1,j_2\in\calM_\calB^2$. Following the similar idea as in \cite{MMD}, as $n\to\infty$, we have 
\begin{align}
\bbE[\mathrm{MMD}^2(Y_{j_2}^n,Y_{j_1}^n)-\mathrm{MMD}^2(Y_i^n,Y_{j_1}^n)]=-\mathrm{MMD}^2(f_\rmN,f_\rmA)    
\end{align}

Define $g_{i,j_1,j_2}(y_i^n,y_{j_1}^n,y_{j_2}^n)=\mathrm{MMD}^2(y_{j_2}^n,y_{j_1}^n)-\mathrm{MMD}^2(y_i^n,y_{j_1}^n)$.
For each $k\in[3n]$, if the $k$-th element of $\{y_i^n,y_{j_1}^n,y_{j_2}^n\}$ is replaced by $\tily$, we use $g_{i,j_1,j_2}(y_i^n,y_{j_1}^n,y_{j_2}^n,k,\tily)$ to denote the corresponding function value. For each $(i,j_1,j_2,k)\in\calB\times\calM_\calB^2\times[3n]$, define
\begin{align}
c_k^{i,j_1,j_2}:=\sup_{\substack{y_i^n,y_{j_1}^n,\\y_{j_2}^n,\tily}}|g_{i,j_1,j_2}(y_i^n,y_{j_1}^n,y_{j_2}^n)-g_{i,j_1,j_2}(y_i^n,y_{j_1}^n,y_{j_2}^n,k,\tily)|.
\end{align}
%

It follows that
\begin{align}
|c_k^{i,j_1,j_2}|&\leq \frac{8K_0}{n},~k\in[3n]
\end{align}
Thus,
\begin{align}\label{frac}
\sum_{k\in[3n]} (c_k^{i,j_1,j_2})^2\leq \frac{192K_0^2}{n}.
\end{align}
Therefore, $\beta_\calB(\Phi(0)|f_\rmN,f_\rmA)$ is upper bounded by
\begin{align}
&\beta_\calB(\Phi_n(0)|f_\rmN,f_\rmA)\\
&\leq 3\sum_{\substack{i\in\calB\\(j_1,j_2)\in\calM_\calB^2\\j_1\neq j_2}}\bbP\{\mathrm{MMD}^2(Y_i^n,Y_{j_1}^n)<\mathrm{MMD}^2(Y_{j_2}^n,Y_{j_1}^n)\}\\
&= 3\sum_{\substack{i\in\calB\\(j_1,j_2)\in\calM_\calB^2\\j_1\neq j_2}}\bbP_\calB\{-\mathrm{MMD}^2(Y_i^n,Y_{j_1}^n)+\mathrm{MMD}^2(Y_{j_2}^n,Y_{j_1}^n)\nn\\
&\qquad-\bbE[\mathrm{MMD}^2(Y_{j_2}^n,Y_{j_1}^n)-\mathrm{MMD}^2(Y_i^n,Y_{j_1}^n)]\nn\\
&\qquad>-\bbE[\mathrm{MMD}^2(Y_{j_2}^n,Y_{j_1}^n)-\mathrm{MMD}^2(Y_i^n,Y_{j_1}^n)]\}\\
\nn&\leq 3s(M-s)^2\\*
&\quad\times \exp\Bigg\{\frac{-2(-\bbE[\mathrm{MMD}^2(Y_{j_2}^n,Y_{j_1}^n)-\mathrm{MMD}^2(Y_i^n,Y_{j_1}^n)])^2}{\sum_{k\in[3n]} (c_k^{i,j_1,j_2})^2}\Bigg\}\\
&=3s(M-s)^2\exp\Bigg\{\frac{-n\mathrm{MMD}^4(f_\rmN,f_\rmA)}{96K_0^2}\Bigg\}\label{error_known2}
\end{align}

Finally, the exponent lower-bound of $\beta_\calB(\Phi_n(0)|f_\rmN,f_\rmA)$ in \eqref{error_known2} is given by \eqref{Exp_known}. 
\begin{align}
\rmE_\calB(\Phi_n(0)|f_\rmN,f_\rmA)
&\geq\frac{\mathrm{MMD}^4(f_\rmN,f_\rmA)}{96K_0^2}
\end{align}

Next, we consider the case of $l>0$
Let $\calE_\mathrm{error}^{(r)}$ be the event of the misclassification during the $r$th-round of iteration, and $\calE_{\calB,\rmA}^{(r)}$ be the event of the re-estimated $\haty_\rmN^n$ being the outlying sequence after $r$ rounds of iteration.
\begin{align}
&\beta_\calB(\Phi(l)|f_\rmN,f_\rmA)\nn\\
&=\beta_\calB(\Phi(l-1)|f_\rmN,f_\rmA)\bbP_\calB\{\calE_{\calB,\rmA}^{(l-1)}|\calE_\mathrm{error}^{(l-1)}\}\nn\\
&+(1-\beta_\calB(\Phi(l-1)|f_\rmN,f_\rmA))\bbP_\calB(\calE_{\calB,3})\\
&\leq \beta_\calB(\Phi(l-1)|f_\rmN,f_\rmA)\bbP_\calB\{\calE_{\calB,\rmA}^{(l-1)}|\calE_\mathrm{error}^{(l-1)}\}+\bbP_\calB(\calE_{\calB,3})\\
&\leq \beta_\calB(\Phi(0)|f_\rmN,f_\rmA)\prod_{r\in[0:l-1]}\bbP_\calB\{\calE_{\calB,\rmA}^{(r)}|\calE_\mathrm{error}^{(r)}\}\nn\\
&+\bbP_\calB\{\calE_{\calB,3}\}\left(1+\sum_{ri\in[0:l-2]}\prod_{r=l-i}^{l-1}\bbP_\calB\{\calE_{\calB,\rmA}^{(r)}|\calE_\mathrm{error}^{(r)}\}\right)\\
&\leq (\bbP_\calB\{\calE_{\calB,1}\}+\bbP_\calB\{\calE_{\calB,2}\})\prod_{r\in[0:l-1]}\bbP_\calB\{\calE_{\calB,\rmA}^{(r)}|\calE_\mathrm{error}^{(r)}\}\nn\\
&+\bbP_\calB(\calE_{\calB,3})\left(1+\sum_{i\in[0:l-1]}\prod_{r=l-1-i}^{l-1}\bbP\{\calE_{\calB,\rmA}^{(r)}|\calE_\mathrm{error}^{(r)}\}\right)\label{betaB1}
\end{align}
We first prove that  $\bbP_\calB\{\calE_{\calB,\rmA}^{(r)}|\calE_\mathrm{error}^{(r)}\}$ is exponential consistent for any $s<\frac{M}{3}$. If $\calE_\mathrm{error}^{(r)}$ happens, for any $\calD\in\calS_s$ and $\calD\neq\calB$,  $\calE_{\calB,\rmA}^{(r)}$ happens when:
\begin{align}
&\exists~i\in \calM_\calD\cap\calB, \exists~j\in \calM_\calD\cap\calM_\calB,\nn\\
&\mathrm{MMD}^2(y_i^n,\bY_{\calM_\calD,i}^n)<\mathrm{MMD}^2(y_{j}^n,\bY_{\calM_\calD,j}^n)
\end{align}
For $n\to \infty$, let $|\calM_\calD\cap\calB|=k$, according to the results in \cite{MMD}, we have  
\begin{align}    
&\bbE[\mathrm{MMD}^2(y_i^n,\bY_{\calM_\calD,i}^n)-\mathrm{MMD}^2(y_{j}^n,\bY_{\calM_\calD,j}^n)]\nn\\
&=\frac{(M-s-2k)((M-s)n-1)}{(M-s-1)((M-s-1)n-1)}\mathrm{MMD}^2(f_\rmN,f_\rmA).
\end{align}
According to McDiarmid’s inequality, we can see that $\bbP_\calB\{\calE_{\calB,\rmA}^{(r)}|\calE_\mathrm{error}^{(r)}\}$ is exponential consistent if $M-s-2k>0$. Therefore, we have $k\leq s<\frac{M-s}{2}$. We let the exponent of $\bbP_\calB\{\calE_{\calB,\rmA}^{(r)}|\calE_\mathrm{error}^{(r)}\}$ be larger than $E$. \eqref{betaB1} can be rewritten as
\begin{align}
&\nn\beta_\calB(\Phi(l)|f_\rmN,f_\rmA)\\
&\leq (\bbP_\calB\{\calE_{\calB,1}\}\nn\\
&+\bbP_\calB\{\calE_{\calB,2}\})e^{-nlE}+\bbP_\calB(\calE_{\calB,3})\left(1+\sum_{i\in[0:l-1]}e^{-(i+1)nE}\right)\\
&< (\bbP_\calB\{\calE_{\calB,1}\}+\bbP\{\calE_{\calB,2}\})e^{-nlE}+\bbP_\calB(\calE_{\calB,3})\left(1+le^{-nE}\right)\\
&<(l+1)e^{-nE}\bbP_\calB\{\calE_{\calB,1}\}+e^{-nlE}\bbP_\calB\{\calE_{\calB,2}\}+\bbP\{\calE_{\calB,3}\}
\end{align}

Recall the definitions of $\calE_{\calB,1}$, $\calE_{\calB,2}$, $\calE_{\calB,3}$ in \eqref{EB1}, \eqref{EB2}, and \eqref{EB3}, respectively. We can find that $\calE_{\calB,2}\subseteq \calE_{\calB,1}$ and $\calE_{\calB,3}\subseteq \calE_{\calB,1}$, thus, $\bbP_\calB\{\calE_{\calB,3}\}\leq \bbP_\calB\{\calE_{\calB,1}\}$, $\bbP_\calB\{\calE_{\calB,2}\}\leq \bbP_\calB\{\calE_{\calB,1}\}$. 
Therefore, for any event $\calE$, let $\mathrm{Exp}(\bbP_\calB\{\calE\})=\lim_{n\to\infty}-\frac{1}{n}\log \bbP_\calB\{\calE\}$ be the probability exponent of $\bbP_\calB\{\calE\}$ 
, we have
\begin{align}
\min\{ \mathrm{Exp}(\bbP_\calB\{\calE_{\calB,2}\}),  \mathrm{Exp}(\bbP_\calB\{\calE_{\calB,3}\})\}\geq\mathrm{Exp}(\bbP_\calB\{\calE_{\calB,1}\}) 
\end{align}

Finally, we have 
\begin{align}
&\rmE_\calB(\Phi(l)|f_\rmN,f_\rmA)\nn\\
&> \min \bigg\{E+\mathrm{Exp}(\bbP_\calB\{\calE_{\calB,1}\}),\nn\\
&\qquad lE+\mathrm{Exp}(\bbP_\calB\{\calE_{\calB,2}\}),\mathrm{Exp}(\bbP_\calB\{\calE_{\calB,3}\})\bigg\}\\
&\geq\min \bigg\{E+\mathrm{Exp}(\bbP_\calB\{\calE_{\calB,1}\})\nn\\
&\qquad,lE+\mathrm{Exp}(\bbP_\calB\{\calE_{\calB,1}\}),\mathrm{Exp}(\bbP_\calB\{\calE_{\calB,3}\})\bigg\}\\
&= \mathrm{Exp}(\bbP_\calB\{\calE_{\calB,1}\})
\end{align}

Recall that 
\begin{align}
&\beta_\calB(\Phi(0)|f_\rmN,f_\rmA) \nn\\
&=\bbP_\calB\{\calE_{\calB,1}\cup \calE_{\calB,2}\}+\bbP_\calB\{\calE_{\calB,1}\cup \calE_{\calB,2})^c\}\bbP\{\calE_{\calB,3}\}\\
&\geq \bbP_\calB\{\calE_{\calB,1}\cup \calE_{\calB,2}\}\\
&\geq\bbP_\calB\{\calE_{\calB,1}\}
\end{align}
we finally derive
\begin{align}
\rmE_\calB(\Phi(0)|f_\rmN,f_\rmA)\leq \mathrm{Exp}(\bbP_\calB\{\calE_{\calB,1}\})<\rmE_\calB(\Phi(l)|f_\rmN,f_\rmA)
\end{align}
The theorem is proved.

\subsection{Proof of Theorem \ref{FL_un}}\label{proof_of_FL_un}

The exponential consistency of $\beta_\calB(\Phi_n|f_\rmN,f_\rmA)$, $\zeta_\calB(\Phi_n|f_\rmN,f_\rmA)$, and $\rmP_{\mathrm{FA}}(\Phi_n|f_\rmN)$ can be established using
techniques similar to those in Theorem \ref{FL}. The
major difference of $\beta_\calB(\Phi_n|f_\rmN,f_\rmA)$ between the proof of Theorem \ref{FL} and Theorem \ref{FL_un}
is that there are two cluster centers in the initialization step and assignment step. Besides, we should consider the impact of $\lambda$ on the exponent bounds of $\zeta_\calB(\Phi_n|f_\rmN,f_\rmA)$ and $\rmP_{\mathrm{FA}}(\Phi(0)|f_\rmN)$.

We first consider the misclassification error exponent.
Due to the structure of the test we know that errors
may occur at two different steps:
\begin{itemize}
\item  The constructed cluster centers  $c_1$ and outlying sequences $c_2$ are actually
generated from the same distribution.
\item The MMD distance of an outlying
sequence is closer to the cluster center of the typical
sequences set, and vice versa.
\item 
\end{itemize}
We use $\calF_{\calB}$ to denote the event that errors occur in the initialization
step. The error event can be decomposed into two
parts. For any $i\in\calB$ and $j\in\calM_\calB$,
define the following two events:
\begin{align}
& \calF_{\calB,1,j}:\exists~i_2\in \calB, \max_{i_1\in\calB}\mathrm{MMD}^2(y_{i_1}^n,y_{i_2}^n)>\mathrm{MMD}^2(y_{i_2}^n,y_j^n)\\
& \calF_{\calB,2,i}:\exists~j_2\in \calM_\calB, \max_{j_1\in\calM_\calB}\mathrm{MMD}^2(y_{j_1}^n,y_{j_2}^n)>\mathrm{MMD}^2(y_{j_2}^n,y_i^n)
\end{align}

Therefore, the error event during the initialization step is given by
\begin{align}
& \calF_{\calB}=\left\{\bigcap_{j\in\calM_\calB}\calF_{\calB,1,j}\right\}\bigcup\left\{\bigcap_{i\in\calB}\calF_{\calB,2,i}\right\}\label{EBun}
\end{align}

We use $\calG_{\calB}$ to denote the event that errors occur at the assignment
step, the error happens when either of the following event happens
\begin{align}
 \calG_{\calB,1}:&~\exists~i_1,i_2\in \calB, \exists~j\in \calM_\calB,\nn\\
&\mathrm{MMD}^2(y_{i_1}^n,y_{i_2}^n)>\mathrm{MMD}^2(y_{i_1}^n,y_j^n)\\
 \calG_{\calB,2}:&~\exists~i\in \calB,\exists~j_1,j_2\in \calM_\calB,\nn\\ &\mathrm{MMD}^2(y_{j_1}^n,y_{j_2}^n)>\mathrm{MMD}^2(y_{j_1}^n,y_i^n)
\end{align}
where $\calG_{\calB,1}$ is the case when the MMD distance of an outlying
sequence is closer to the cluster center of the typical
sequences set, and $\calG_{\calB,2}$ is the case when the MMD distance of a typical
sequence is closer to the cluster center of the outlying
sequences set. Thus, we have $\calG_\calB=\calG_{\calB,1}\cup \calG_{\calB,2}$
Therefore, $\beta_\calB$ can be upper bounded by
\begin{align}
\beta_\calB(\Phi_n|f_\rmN,f_\rmA)
&=\bbP_\calB\{\calF_{\calB}\}+\bbP_\calB\{\calF_{\calB}^c\cap\calG_\calB\}\\
&\leq \bbP_\calB\{\calF_{\calB}\}+\bbP_\calB\{\calG_{\calB,1}\}+\bbP\{\calG_{\calB,2}\}\label{beta_error_unknown}
\end{align}
To compute \eqref{beta_error_unknown}, we first derive the upper bound of $\bbP_\calB\{\calF_{\calB}\}$, which is given by
\begin{align}
&\bbP_\calB\{\calF_{\calB}\}\nn\\
&\leq \min_{j\in\calM_\calB}\bbP_\calB\{\calF_{\calB,1,j}\}+\min_{i\in\calB}\bbP_\calB\{\calF_{\calB,2,i}\}\\
&\leq \min_{j\in\calM_\calB}\sum_{i_2\in\calB}\bbP_\calB\{\max_{i_1\in\calB}\mathrm{MMD}^2(y_{i_1}^n,y_{i_2}^n)>\mathrm{MMD}^2(y_{i_2}^n,y_j^n)\}\nn\\
&+\min_{i\in\calB}\sum_{j_2\in\calM_\calB}\bbP_\calB\{\max_{j_1\in\calM_\calB}\mathrm{MMD}^2(y_{j_1}^n,y_{j_2}^n)>\mathrm{MMD}^2(y_{j_2}^n,y_i^n)\}\\
&\leq \min_{j\in\calM_\calB}\sum_{i_2\in\calB}\bbP_\calB\{\exists ~i_1\in\calB, \mathrm{MMD}^2(y_{i_1}^n,y_{i_2}^n)>\mathrm{MMD}^2(y_{i_2}^n,y_j^n)\}\nn\\
&+\min_{i\in\calB}\sum_{j_2\in\calM_\calB}\bbP_\calB\{\exists~j_1\in\calM_\calB,\mathrm{MMD}^2(y_{j_1}^n,y_{j_2}^n)>\mathrm{MMD}^2(y_{j_2}^n,y_i^n)\}\\
&\leq  \min_{j\in\calM_\calB}\sum_{i_1,i_2\in\calB^2}\bbP_\calB\{\mathrm{MMD}^2(y_{i_1}^n,y_{i_2}^n)>\mathrm{MMD}^2(y_{i_2}^n,y_j^n)\}\nn\\
&+\min_{i\in\calB}\sum_{j_1,j_2\in\calM_\calB^2}\bbP_\calB\{\mathrm{MMD}^2(y_{j_1}^n,y_{j_2}^n)>\mathrm{MMD}^2(y_{j_2}^n,y_i^n)\}
\end{align}
By following the similar idea of \eqref{error_known}, we have
\begin{align}
\bbP_\calB\{\calF_{\calB}\}&\leq (s^2+(M-s)^2)\exp\Bigg\{-\frac{n\mathrm{MMD}^4(f_\rmN,f_\rmA)}{96K_0^2}\Bigg\}\label{EBun2}
\end{align}

We can see that the lower bound of the exponent of $\bbP_\calB\{\calF_{\calB}\}$ is the same as $\bbP_\calB\{\calG_{\calB,1}\}+\bbP_\calB\{\calG_{\calB,2}\}$. 
Therefore, the exponents of $\beta_\calB(\Phi_n|f_\rmN,f_\rmA)$ is lower bounded by
\begin{align}
\rmE_{\beta_\calB}(\Phi_n|f_\rmN,f_\rmA)\geq\frac{\mathrm{MMD}^4(f_\rmN,f_\rmA)}{96K_0^2}\label{mis_error_exponent_unknown2}
\end{align}
\eqref{mis_error_exponent_unknown} shows that the achievable misclassification error exponent in the known case is smaller than that of known case, which is the penalty of not knowing the number of outliers.

By following the similar idea, the false reject probability and false alarm probability, $\zeta_\calB(\Phi_n|f_\rmN,f_\rmA)$, and $\rmP_\mathrm{FA}(\Phi_n|f_\rmN,f_\rmA)$ can be upper bounded by $\bbP_\calB\{\calF_{\calB,3}\}$ and $\bbP\{\calF_4\}$. Here, $\calF_{\calB,3}$ and $\calF_4$ which are given by:
\begin{align}
& \calF_{\calB,3}:~\forall~i\in\calB,j\in\calM_\calB,\mathrm{MMD}^2(y_i^n,y_j^n)<\lambda\\
&  \calF_4:~\exists~i,j\in [M]^2,\mathrm{MMD}^2(y_i^n,y_j^n)\geq\lambda
\end{align}
Therefore, we have 
\begin{align}
&\zeta_\calB(\Phi_n|f_\rmN,f_\rmA)\leq \bbP_\calB\{\calF_{\calB,3}\}\\
&=\bbP_\calB\{\forall~i\in\calB,j\in\calM_\calB,\mathrm{MMD}^2(y_i^n,y_j^n)<\lambda\}\\
&\leq \min_{i\in\calB,j\in\calM_\calB }\bbP_\calB\{\mathrm{MMD}^2(y_i^n,y_j^n)<\lambda\}\\
&\leq \exp\Bigg\{\frac{-n(\mathrm{MMD}^2(f_\rmN,f_\rmA)-\lambda)^2}{64K_0^2}\Bigg\}\label{zeta_unknown}
\end{align}
Similarly, 
\begin{align}
\rmP_\mathrm{FA}(\Phi_n|f_\rmN,f_\rmA)&\leq\bbP_\rmr\{\calF_4\}\\
&=\bbP_\rmr\{\exists~i,j\in [M]^2,\mathrm{MMD}^2(y_i^n,y_j^n)\geq\lambda\}\\
&\leq \sum_{i,j\in [M]^2}\bbP_\rmr\{\mathrm{MMD}^2(y_i^n,y_j^n)\geq\lambda\}\label{FA_un1}\\
&\leq \sum_{i,j\in [M]^2} \exp\left\{\frac{-n\lambda^2}{64K_0^2}\right\}\label{FA_unknown}
\end{align}
Therefore, we can bound the exponents of $\zeta_\calB(\Phi_n|f_\rmN,f_\rmA)$ and $\rmP_{\mathrm{FA}}(\Phi_n|f_\rmN,f_\rmA)$ as
\begin{align}
&\rmE_{\zeta_\calB}(\Phi_n|f_\rmN,f_\rmA)\geq \frac{(\mathrm{MMD}^2(f_\rmN,f_\rmA)-\lambda)^2}{64K_0^2}\\
& \rmE_{\mathrm{FA}}(\Phi_n|f_\rmN)\geq \frac{\lambda^2}{64K_0^2}
\end{align}

\end{document}